\documentclass[journal,twocolumn]{IEEEtran}
\usepackage{xcolor}
\usepackage{mathtools}
\usepackage{epsfig,makeidx,color,subfigure}
\usepackage{amsmath,amssymb,bbm,enumitem}
\usepackage{cite,graphicx,lipsum}
\usepackage[switch,pagewise]{lineno}
\usepackage{tabularx}
\usepackage{hyperref}
\hypersetup{
        colorlinks = true,
        citecolor = red,
        breaklinks = true,
        linktocpage=true,
}
\usepackage{url}

\def\be{ \begin{equation} }
\def\ee{ \end{equation} }
\def\bea{ \begin{eqnarray} }
\def\eea{ \end{eqnarray} }

\def\b0{{\bf 0}}

\def\cL{{\cal L}}

\ifCLASSOPTIONonecolumn
\else

\fi

\begin{document}

\title{Spectrum Sharing through Marketplaces for O-RAN based Non-Terrestrial and Terrestrial Networks}
\author{Jinho Choi, Bohai Li, Bassel Al Homssi, Jihong Park, and Seung-Lyun Kim
}

\maketitle
\begin{abstract}
Non-terrestrial networks (NTNs), including low Earth orbit (LEO) satellites, are expected to play a pivotal role in achieving global coverage for Internet-of-Things (IoT) applications in sixth-generation (6G) systems. Although specific frequency bands have been identified for satellite use in NTNs, persistent challenges arise due to the limited availability of spectrum resources. The coexistence of multiple systems, including terrestrial networks (TNs), sharing these frequencies, presents a technically challenging yet feasible solution. Furthermore, the effective management and regulation of such coexistence should be under the purview of regional authorities. To facilitate efficient spectrum sharing among various systems, including NTNs and TNs, adopting open architectures is desirable, allowing for the seamless exchange of key information for spectrum sharing. Therefore, it is essential to consider open radio access networks (O-RAN) for future NTNs and TNs. In addition to O-RAN, the establishment of spectrum marketplaces, enabling different operators to trade their spectrum and dynamic resource allocation information, is necessary. In this article, we highlight the role of spectrum marketplaces and discuss a few examples.
\end{abstract}


\ifCLASSOPTIONonecolumn
\baselineskip 28pt
\fi

\section{Introduction}

True global coverage is the zenith of connectivity and promises a world where opportunities are not bound by geography, where every individual, irrespective of their location, can participate in the global economy. In the pursuit of achieving true global coverage, next-generation wireless communication networks are envisioned to establish non-terrestrial networks (NTNs) \cite{Giordani21} \cite{Azari22}. These NTNs represent a paradigm shift in the field of telecommunications, as they incorporate a diverse range of communication systems beyond traditional terrestrial infrastructure such as cellular networks. NTNs encompass a multitude of technologies, including low-earth orbit (LEO) satellites, unmanned aerial vehicles (UAVs), and other non-terrestrial platforms, all working together to provide seamless, global connectivity. By extending the reach of wireless communication to previously underserved or hard-to-reach areas, NTNs aim to bridge the digital divide and unlock new possibilities for communication and connectivity on a global scale.

While NTNs can promise global connectivity, the allocation of radio spectrum resources for NTNs is a critical challenge due to the limited and increasingly crowded spectrum. As a viable solution, spectrum sharing between NTNs and terrestrial networks (TNs) emerges \cite{Martikainen2023}.

Open Radio Access Network  (O-RAN) stands at the forefront of a transformative shift in the telecommunications industry 
\cite{Bonati20} \cite{Polese23}. It represents a departure from traditional, closed network architectures by offering a flexible and open framework that allows mobile network operators to mix and match equipment from various vendors while ensuring seamless interoperability through RAN Intelligent Controllers (RIC). In this rapidly evolving landscape, O-RAN can play a pivotal role in enabling efficient spectrum utilization by facilitating sharing between different networks, including NTNs.

With O-RAN, there are two primary approaches for efficient spectrum utilization and a hybrid approach that combines elements of both as follows. 
\begin{itemize}
\item Cognitive Radio Systems: In this approach, a primary system is allocated a licensed spectrum. Secondary systems aim to utilize spectrum ``holes" that are not actively used by the primary system \cite{Liang11}. To operate effectively, secondary systems must possess information about the spectrum occupancy of the primary system. This information can be obtained through spectrum sensing or other methods. A near-real-time RIC can be used to provide the resource allocation information of a primary system to secondary systems.

\item Spectrum Sharing Systems: Multiple systems each have their own allocated bandwidth. While static spectrum sharing can be achieved through spectrum utilization scheduling via a non-real-time RIC, implementing dynamic spectrum sharing can be challenging. The challenge arises because real-time resource allocation information needs to be shared among different systems. Note that, as discussed in \cite{Baldesi22}, it would be possible to perform spectrum sharing using a near-real-time RIC with sensing capability. However, this approach would be more closely aligned with the above-mentioned approach based on cognitive radio, while our spectrum sharing systems primarily rely on scheduling to allocate shared spectrum.

\item Hybrid Approach: In spectrum sharing systems, the aggregated bandwidth is divided into multiple resource blocks, and groups of blocks are allocated based on demand in a scheduling process that operates at a relatively slow time scale through a non-real-time RIC. The utilization of each resource block is system-dependent, and it is possible that a resource block may not be fully utilized by the system that initially receives it. In such cases, other systems, functioning as secondary systems, can opportunistically exploit these under-utilized resource blocks using faster time-scale operations enabled by spectrum sensing or a near-real-time RIC. This hybrid approach combines the primary resource allocation process with secondary, opportunistic spectrum access.
\end{itemize}
While secondary systems can reap advantages from cognitive systems, primary systems, unfortunately, typically do not enjoy such benefits. In contrast, spectrum sharing systems provide gains for all participating systems, serving as a motivating factor for primary systems to engage in the marketplace.

While O-RAN can facilitate spectrum sharing through marketplaces across different networks using one of the above approaches, one of its most significant advantages becomes evident when applied to Internet of Things (IoT) networks that span vast regions, transcending international borders. O-RAN based spectrum sharing empowers IoT networks to efficiently utilize frequencies designated for other networks, even in cases where the IoT networks do not own licensed bands. This unique capability allows networks to offer connectivity services over extensive areas, including multiple countries, by dynamically accessing under-utilized spectrum. In this article, we discuss O-RAN-based spectrum sharing with NTNs and TNs, highlighting the role of spectrum marketplaces.

The rest of this article is as follows. We introduce O-RAN and marketplaces for spectrum sharing in Section~\ref{S:ORAN_SM}. Two use cases are discussed in Sections~\ref{S:UC1} and \ref{S:UC2} to demonstrate how O-RAN and spectrum marketplaces can help share the spectrum between different networks. We elaborate on the gains of spectrum sharing in Section~\ref{S:CSS}. A few challenges are addressed in Section~\ref{S:Challenges}. Finally, we conclude the article with remarks in Section~\ref{S:Conc}.



\section{Open-RAN and Spectrum Marketplaces}    \label{S:ORAN_SM}


To achieve efficient spectrum sharing, O-RAN is essential. However, depending solely on O-RAN may confine spectrum sharing to long-term mutual agreements among a limited number of participating networks. To enhance flexibility and encourage broader participation, the concept of spectrum marketplaces becomes crucial. These marketplaces serve as dynamic platforms that incentivize and facilitate the involvement of numerous networks, fostering a more inclusive and adaptable allocation of spectrum resources. In this section, we provide a brief overview of the roles played by both O-RAN and spectrum marketplaces in utilizing the scarce spectrum resource.

\subsection{Open-RAN}

O-RAN introduces a groundbreaking approach to RAN architecture, which is segmented into three essential building blocks: the Radio Unit (RU), the Distributed Unit (DU), and the Centralized Unit (CU) \cite{Polese23}. These building blocks play a crucial role in reshaping the structure of mobile networks and offer various advantages. The RU, located near or integrated into antennas, manages radio signal transmission, reception, amplification, and digitization. On the other hand, the DU and CU, strategically positioned within the network, oversee the computational aspects of the base station. O-RAN's white-box approach is a key innovation, opening up protocols and interfaces between these building blocks, allowing for enhanced flexibility and vendor diversity in network deployment. 


\subsubsection*{Spectrum Sharing and Coordination}
O-RAN's open architecture not only enables practical spectrum resource sharing but also lays the foundation for seamless coordination among diverse systems. The inherent flexibility in spectrum occupancy is further enhanced through O-RAN's open interfaces, creating a more versatile network environment. Additionally, the introduction of an intelligent controller emerges as a powerful toolset for efficient spectrum resource management, unlocking new possibilities for coordinated spectrum utilization.

\subsubsection*{Example of Collaborative Spectrum Sharing}
In \cite{Smith21}, spectrum sharing approaches between heterogeneous systems enabled by O-RAN are thoroughly examined, with a strong emphasis on the potential for collaborative use cases. A standout example involves the synergy between government LEO satellites and 5G uplinks, demonstrating O-RAN's remarkable capabilities in bridging these distinct systems. Furthermore, the research explores the prospects of government-managed 5G O-RAN systems and the intriguing concept of government-to-government spectrum sharing.

\subsection{Spectrum Marketplaces}

While O-RAN's open architecture sets the stage for enhanced spectrum sharing and intelligent coordination among diverse systems, the adoption of O-RAN ultimately rests in the hands of network operators. As more operators embrace O-RAN technology, it opens up the possibility of creating dynamic spectrum marketplaces, which can have far-reaching implications for the telecommunications industry.

In marketpalaces, there are key elements as follows.
\begin{itemize}
    \item Empowering Network Operators as Sellers and Buyers: With network operators actively incorporating O-RAN into their infrastructure, a spectrum marketplace can take shape, allowing each operator to assume the roles of both a spectrum seller and buyer. This dual capacity empowers operators to optimize their spectrum utilization, making more efficient use of their allocated frequencies.
    \item Spectrum Brokers: In addition to network operators, the spectrum marketplace may also introduce the role of spectrum brokers. These specialized entities serve as intermediaries, facilitating spectrum sharing and coalitions among various stakeholders. The involvement of brokers streamlines the process of spectrum allocation, negotiation, and coordination, fostering a more agile and responsive telecommunications landscape.
\end{itemize}

The emergence of spectrum marketplaces signifies a shift towards collaborative and efficient spectrum management. Operators can trade or lease their excess spectrum to others in need, promoting resource optimization and reducing spectrum wastage. This cooperative approach benefits not only the participating operators but also the end-users who experience improved network performance and expanded service offerings.

Spectrum marketplaces can extend their reach beyond traditional network operators. Companies seeking to deploy private networks, IoT services, or other wireless solutions can also participate as buyers in these marketplaces. This diversity of participants enriches the spectrum marketplace ecosystem, fostering innovation and resource-sharing across various sectors.

\section{Use Case 1: Cognitive Satellite Systems}   \label{S:UC1}

In this section, we discuss frequency reuse based on cognitive radio, whereby, a primary and a secondary system coexist, with the secondary system aiming to capitalize on the unutilized spectrum holes within the bandwidth allocated to the primary system. This highlights the potential for collaborative spectrum sharing between different systems, enabling the secondary system to efficiently harness under-utilized spectrum resources. The unique challenge that arises is that the secondary system must navigate the intricacies of resource allocation with limited information about the primary system's usage patterns which are often highly dynamic. Hence, the secondary system must employ sophisticated techniques such as spectrum sensing and dynamic spectrum management to adapt and optimize its resource allocation in real-time while avoiding interfering with primary system transmissions. Alternatively, an open architecture approach can be used to effectively provide the necessary information, which can be exchanged through spectrum marketplaces. We will explore this approach further after introducing the co-existing LEO and GEO scenario.

\subsection{Co-Existing LEO and GEO}

\begin{figure}
\centering
\includegraphics[width=0.9\linewidth]{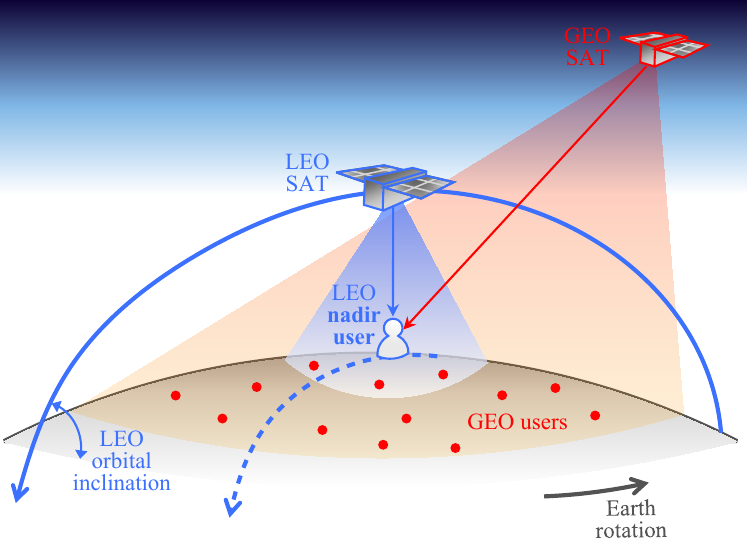} 
\vspace{-0.2cm}
\caption{Integrated GEO-LEO SAT networks where GEO SATs are primary and LEO SATs are secondary.}
\label{Fig:Fig1}
\vspace{-0.37cm}
\end{figure}

Consider GEO and LEO satellites that overlap in their serving region, as illustrated in Fig.~\ref{Fig:Fig1}, where the secondary system is considered to be the LEO satellites and the primary system is the GEO satellites. A GEO operator can provide resource allocation data to a LEO operator, enabling LEO satellites to utilize unallocated resources within their footprints. This practice can be extended to multiple GEO operators with dedicated spectrum resources available for reuse by other LEO operators. Since the footprint of LEO is much smaller than that of the GEO, the downlink signals from a LEO satellite may not interfere with most users supported by the GEO satellite who are located outside of the LEO's footprint. Thus, as long as the LEO satellite avoids using the allocated spectrum for the GEO users within the overlapping region, which would be a fraction of the total allocated bandwidth to the GEO, the LEO satellite can freely use the spectrum. 

\begin{figure}
\centering
\includegraphics[width=0.9\linewidth]{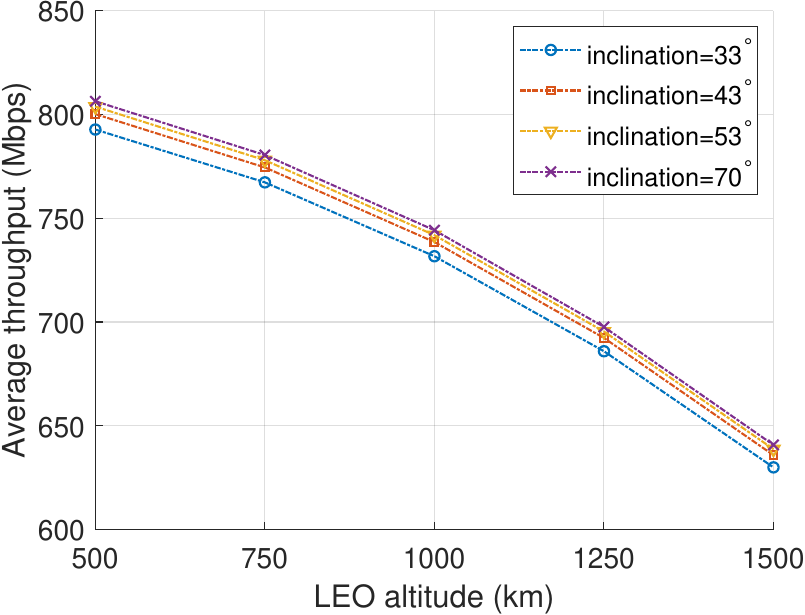} 
\vspace{-0.2cm}
\caption{Average throughput of the LEO system as a function of altitude.}
\label{Fig:Throughput}
\vspace{-0.37cm}
\end{figure}

Fig.~\ref{Fig:Throughput} depicts the LEO system's average throughput as a function of LEO altitude when the L-band (1-2 GHz) spectrum, assumed to be 200 MHz at a 2 GHz frequency, is shared with the GEO system. The results demonstrate that throughput decreases as the LEO's altitude increases, while the impact of the LEO's inclination on the throughput is insignificant. This reduction in throughput is due to the expanding footprint of the LEO system as its altitude rises, providing fewer opportunities for LEO to reuse the GEO's spectrum. 

While Fig.~\ref{Fig:Throughput} illustrates effective frequency re-use in the L-band, the Ka band (26.5–40 GHz) can  also be shared by both GEO and LEO satellites, offering substantial bandwidths and high data transmission rates. Presently, the Ka band is experiencing increased congestion due to its widespread use in satellite communications. Numerous satellite systems, positioned in different orbits, now operate within this spectrum range, each with distinct frequency allocations. For example, the Inmarsat I-5, a GEO system, relies on this band, while SpaceX's Starlink constellation, a LEO network, also utilizes the Ka band. To address the congestion issue, a cognitive radio scenario can be considered, with Inmarsat as the primary system and Starlink as the secondary system. In this configuration, satellites within the Starlink constellation can intelligently utilize unoccupied spectrum segments. When we extend this analysis to a full-sized Starlink constellation, the extensive coverage provided by the large number of LEO satellites, often overlapping with Inmarsat satellites, becomes significant and, in most cases, surpasses it.

As a result, an intelligent mechanism that leverages spatiotemporal diversity is extremely crucial to maintain the integrity of the primary system's transmissions. For example, Starlink's satellites sharing the coverage area with Inmarsat can be organized into clusters, where each cluster is assigned different frequency slots within the band. While this reduces the effects of interference, it may lead to reduced data rates and increased spectrum segmentation. Alternatively, each cluster can intelligently assign orthogonal pilots to its satellites and segment the transmissions temporally, similar to random access networks. The clustering must be dynamic, with every new LEO satellite within the GEO coverage joining the cluster and departing when it leaves. This is necessary because LEO satellites are in constant orbit, while GEO satellites remain static in space.

\subsection{Opportunistic Updating Models for Intelligent IoT}

The re-use of channels by LEO satellites presents exciting opportunities for various intelligent IoT applications. Beyond their ability to collect data, these intelligent IoT sensors and devices hold the potential to locally process data, enabling on-site decision-making through the application of AI/ML algorithms. This signifies that these devices are not only collect environmental data (or measurements) but can also perform real-time, context-aware analysis, facilitating prompt decision-making.

However, effective AI/ML model training often necessitates a substantial amount of data and considerable computing power. This can pose a challenge for IoT devices that are operated in remote or resource-constrained areas, which may lack access to such extensive computational resources. In such scenarios, a pragmatic solution involves training AI models within cloud-based environments, where copious data and computational resources are readily available. Subsequently, these trained models can be seamlessly downloaded to IoT devices, empowering them with advanced AI capabilities.

As previously discussed, LEO satellites possess the capability to efficiently re-use spectrum resources initially allocated to GEO satellites. They can effectively support a multitude of intelligent IoT devices by updating their AI models. Capitalizing on this capacity, LEO satellites can serve as a conduit for disseminating trained model parameters to IoT devices spanning vast geographical areas. This collaborative integration of LEO satellite technology and intelligent IoT applications ensures that even devices in the most remote locations can leverage cutting-edge AI advancements, bolstering their functionality and enabling more intelligent and autonomous decision-making processes.

\subsection{Roles of Marketplaces}

As mentioned earlier, the LEO system as a secondary system needs to have the radio resource allocation of the GEO system. Based on an agreement between GEO and LEO operators, the GEO operator can provide this information to the LEO operator. Alternatively, they can utilize spectrum marketplaces. In particular, the notion of satellite spectrum markets can introduce a dynamic framework wherein GEO operators, acting as sellers, collaborate with LEO operators, who function as buyers, to efficiently allocate and share spectrum resources. In this context, a GEO operator not only provides valuable resource allocation information to a LEO operator but also opens up opportunities for LEO satellites to access unallocated spectrum resources within their coverage footprints.


Note that O-RAN plays a pivotal role in facilitating the exchange of spectrum allocation information through well-established standards and protocols governing the format and structure of exchanged data. This standardization streamlines the process, ensuring that critical information regarding spectrum allocation, utilization, and availability adheres to predefined formats. These standards not only promote seamless information exchange between GEO and LEO operators but also enable efficient communication between operators and the spectrum marketplace.

\section{Use Case 2: Cognitive NTNs}   \label{S:UC2}

In this section, we discuss cognitive NTNs, in which satellite systems function as secondary systems to opportunistically utilize the spectrum allocated to TNs as primary systems. These NTNs can facilitate IoT services in rural areas by utilizing the TNs' allocated spectrum.

\subsection{Co-Existing Terrestrial and Satellite Systems}
A significant amount of bandwidth has been allocated to TNs, such as LTE and 5G, to meet the growing demand for wireless connectivity. While the frequencies assigned to TNs are actively utilized in densely populated areas, such as urban centers, they often remain under-utilized in sparsely populated regions, including remote areas like oceans and deserts. Furthermore, in various satellite-based IoT applications, sensors and devices are strategically positioned to monitor environments with limited human activity, such as remote wildlife tracking or environmental sensing in rural areas \cite{Sancti16}. As a result, significant portions of the frequency spectrum allocated to TNs remain untapped in these low-human-population regions.

\begin{figure}[!t]
\begin{center} \hspace{-1.5em} \subfigure[Trajectory of LEO satellite]{\label{fig 0 ax}\includegraphics[width=0.45\columnwidth]{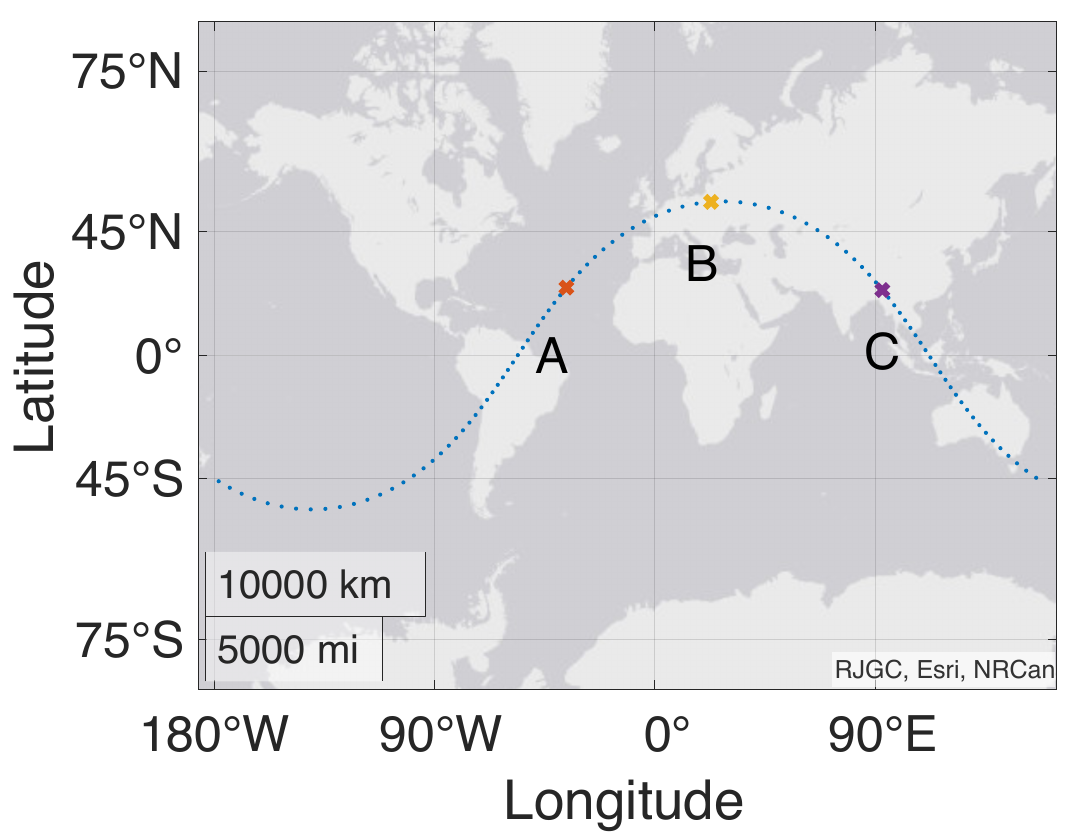}}
\subfigure[SINR over revolution time]{\label{fig 0 bx}\includegraphics[width=0.45\columnwidth]{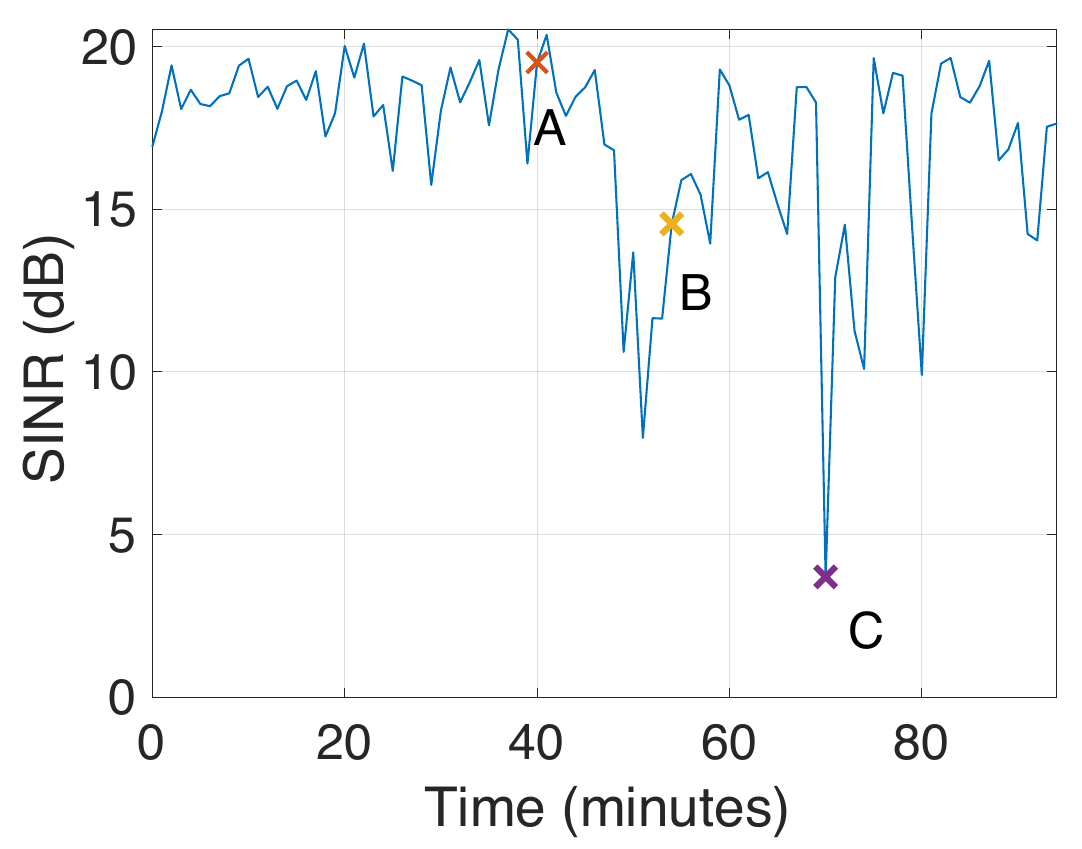}} 
\end{center}
\caption{Frequence-reuse by satellite IoT services: (a) a trajectory of LEO satellite; (b) the SINR over revolution time.}
     \label{Fig:LEO_SINR}
\end{figure}

The under-utilized spectrum resource represents an opportunity for satellite IoT to extend its coverage and support a variety of sensors and devices situated in regions with sparse human populations, thus enhancing the reach and efficiency of satellite IoT services. This reallocation of frequencies, within the framework of NTNs, holds the potential to bridge the digital divide, ensuring that even remote and underserved areas can benefit from the capabilities of satellite IoT solutions. Fig.~\ref{Fig:LEO_SINR} (a) shows the trajectory of the LEO satellite, capable of receiving signals from IoT devices through the spectrum allocated to TNs (e.g., 5G systems operating at 28 GHz). When the LEO satellite passes over highly populated areas (such as point C), high interference from cellular users is expected, and the resulting signal-to-interference-plus-noise ratio (SINR) becomes lower, as illustrated in Fig.~\ref{Fig:LEO_SINR} (b). It is also noted that a high SINR can be achieved when the LEO satellite passes over unpopulated areas, such as oceans (e.g., point A). This indicates that IoT devices can reuse the allocated spectrum to TNs when deployed in remote areas for the application of the Internet of Remote Things.


Through spectrum marketplaces, under-utilized spectrum from TNs can be shared with satellite IoT applications, enabling more efficient resource utilization. As depicted in Fig.~\ref{Fig:LEO_SINR} (b), TNs can permit IoT devices to use their spectrum, particularly when operating in remote areas. In this scenario, various operators from both TNs and NTNs can actively participate in the spectrum marketplace. IoT devices and sensors can seamlessly communicate with LEO satellites, capitalizing on this shared spectrum.


\subsection{Spectrum Sensing by O-RAN based TNs}

Within the O-RAN architecture, RU stands as an independent unit, allowing it to perform the crucial task of spectrum sensing \cite{Baldesi22}. By receiving and analyzing RF signals within their designated frequency bands, RUs hold the capability to provide valuable spectrum sensing information. Consequently, there is a feasible option to gather spectrum sensing data from all RUs integrated into the O-RAN based TNs through the open interface.

The aggregation of this spectrum sensing information can give rise to what is commonly referred to as the Radio Environmental Map (REM), which might be a component of a spectrum management system in O-RAN. The REM serves as a comprehensive and dynamic repository of information concerning the utilization of allocated frequencies within TNs. It offers a real-time overview of the spectral landscape, enabling efficient and informed decisions regarding spectrum allocation and management. Such detailed and up-to-date REMs play a pivotal role in enhancing spectrum utilization, reducing interference, and ensuring the seamless coexistence of various systems within the O-RAN framework.

Furthermore, the REM server plays a pivotal role within the spectrum marketplace, acting as a crucial information provider for satellite IoT service providers as an example. Its function is twofold: first, it aggregates real-time spectrum sensing information from RUs within the O-RAN based TNs, creating a dynamic REM that reflects the ever-changing spectrum environment. This REM data becomes a valuable asset, as it offers satellite IoT service providers insights into available frequencies, interference patterns, and spectrum utilization in real-time, allowing them to make informed decisions for efficient resource allocation. Second, the REM server seamlessly integrates within the spectrum marketplace, serving as an essential element for cooperative spectrum sharing. It provides spectrum intelligence that empowers satellite IoT service providers to identify unoccupied or under-utilized frequencies allocated to TNs, enabling them to harness these resources effectively for their IoT applications. The REM server, therefore, serves as a bridge between terrestrial and satellite-based services, facilitating improved spectrum management, efficient usage, and cooperative arrangements within O-RAN spectrum marketplaces, ultimately benefiting underserved regions with under-utilized spectrum of TNs and advancing the reach of satellite IoT solutions.

\section{Cooperative Spectrum Sharing}  \label{S:CSS}

In the previous two sections, we studied cognitive systems for efficient spectrum utilization, where primary systems (namely GEO systems and TNs) cannot improve their spectrum utilization although they could achieve financial benefits by selling their dynamic spectrum allocation information to LEO operators.In this section, we shift our attention to spectrum sharing, where each system possesses its own bandwidth for sharing. We commence by demonstrating the advantages of spectrum sharing using a cooperative game model. Then, we discuss the application to NTNs.

\subsection{Cooperative O-RAN Game} \label{SS:game}

It is essential to demonstrate the advantages of spectrum sharing between different systems using O-RAN, which can encourage operators to agree to share their bandwidth for each other's services through a marketplace. This can be achieved by employing a cooperative game model \cite{Peleg07}.

Suppose there are $N$ players, each of whom is an operator. 
While each player can operate independently without forming a coalition, a group of operators, denoted by $S$, can agree to share their bandwidth, referred to as a coalition. 
The characteristic function or payoff is a function of coalition, $S$, i.e., $v(S)$, which the gain resulting from a coalition $S$.
Using the notion of the effective bandwidth \cite{Kelly_Yudovina}, for a givenquality-of-service (QoS) exponent, we can determine the payoff as the maximum average connection rate.

\begin{figure}[!t]
\begin{center} \hspace{-1.5em} \subfigure[Payoff versus $\gamma$]{\label{fig 1 ax}\includegraphics[width=0.45\columnwidth]{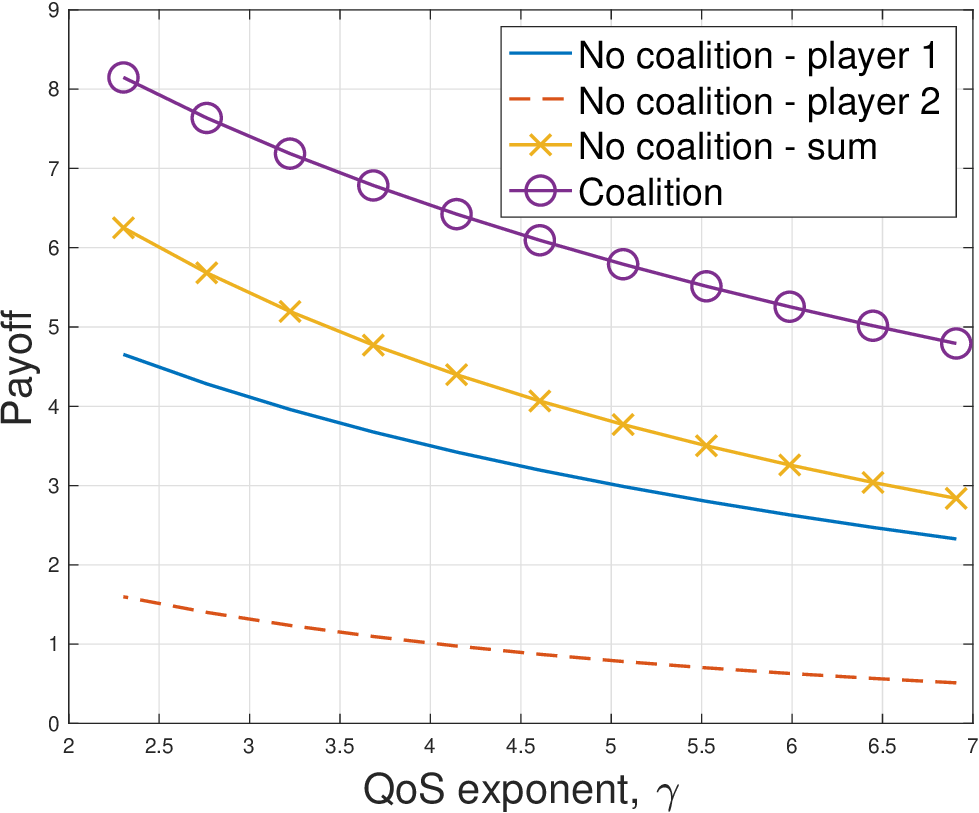}}
\subfigure[Sum of payoff versus $\gamma$]{\label{fig 1 bx}\includegraphics[width=0.45\columnwidth]{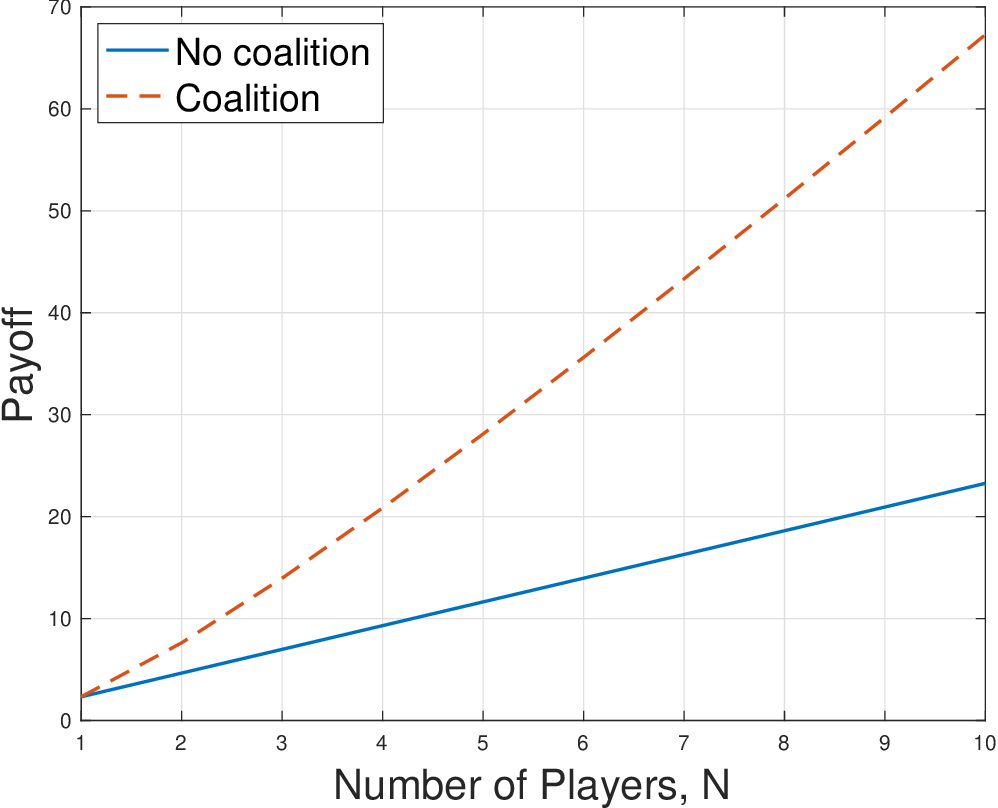}} 
\end{center}
\caption{Payoff of cooperative O-RAN game: (a) payoff versus QoS exponent (with $B_1 = 10$ and $B_2 = 5$); (b) sum of payof versus $N$ (with $e^{-\gamma} = 10^{-3}$ and $B_n = 10$ for all $n$).}
     \label{Fig:plt_game}
\end{figure}

Fig.~\ref{Fig:plt_game} illustrates the payoffs with and without coalition. In Fig.~\ref{Fig:plt_game} (a), we explore a scenario involving two operators with available bandwidths denoted as $B_1 = 10$ and $B_2 = 5$, where $B_n$ represents the available bandwidth for player $n$. The figure presents the payoffs as the QoS exponent, denoted by $\beta$, varies. Here, in this context, $e^{-\beta}$ becomes the outage rate, which is the probability that an operator cannot meet pre-defined performance criteria in terms of connectivity.
We can see that the payoff decreases with the QoS exponent as a higher QoS is expected, while the payoff with coalition is higher than that without coalition. 
Furthermore, in Fig.~\ref{Fig:plt_game} (b), we can see that the sum of payoff as a function of the number of players in a coalition when $e^{-\gamma} = 10^{-3}$ and $B_n = 10$ for all $n$. 
While the sum of payoffs for independent players increases linearly with $N$, the payoff of a coalition increases rapidly and exceeds that of non-coalition scenarios. As a result, operators in a coalition experience higher payoffs as the coalition size increases.

The ability of O-RAN to provide standardized and interoperable frameworks motivates operators to efficiently share bandwidth and optimize their network resources. O-RAN's open architecture fosters seamless communication and coordination among different operators and network elements, simplifying bandwidth sharing and resource allocation. Operators can establish a dynamic and adaptable network ecosystem, flexibly allocating resources to meet changing user and application demands via spectrum marketplaces.

\subsection{Cooperative NTNs and TNs}

A NTN  can be envisioned as a sophisticated integrated network comprising diverse systems, each allocated its dedicated bandwidth resources. For instance, LEO systems, UAV networks, and TNs possess their individual spectrum allocations with a degree of interoperability. However, as demonstrated above, the power of O-RAN becomes evident in its ability to enable these diverse systems to efficiently share their bandwidth through open interfaces and protocols. 

For instance, in Japan, since satellite and 5G systems operate on the same frequencies in the C-band, a cooperative approach has been proposed to mitigate the interference from 5G base stations when receiving satellite signals \cite{Fujii23}. However, implementing interference cancellation would require sophisticated hardware, making it costly and infeasible. Instead, O-RAN can serve as a solution for both systems to share frequencies in the C-band, thereby increasing effective bandwidth, as discussed in Subsection~\ref{SS:game}.

Another notable example pertains to millimeter-wave (mmWave) usage. In the context of 5G networks, Integrated Access and Backhaul (IAB) technology permits the utilization of cellular frequencies for both backhaul and access \cite{Polese20}, leading to reduced deployment costs. mmWave frequencies are commonly considered for this purpose \cite{Li17}. Furthermore, mmWave frequencies are being explored for UAV networks \cite{Xiao22}, and in addition to this, LEO satellites operate within the Ka-band, which shares spectral characteristics with mmWave frequencies. Consequently,  these three distinct systems, namely 5G wireless backhaul, UAV networks, and LEO satellites, may operate on frequencies that are in close proximity to each other, presenting an enticing opportunity for sharing bandwidth resources through spectrum marketplaces and fostering the development of cooperative NTNs and TNs via O-RAN.

\section{Challenges}    \label{S:Challenges}

In this section, we discuss various challenges when O-RAN based spectrum sharing is used for NTNs and TNs.

\subsubsection*{NTN Spectrum Sharing}
A key challenge in spectrum sharing pertains to ensuring that all participating parties adhere to their agreements, a critical element in averting unintended interference. This challenge assumes even greater significance in the context of NTNs, where the overarching objective is to establish extensive coverage across multiple nations. While spectrum marketplaces serve as instrumental platforms for the buying and selling of spectrum resources, international regulatory oversight is imperative to guarantee that all parties involved comply with their commitments regarding spectrum sharing and utilization. This challenge is less pronounced when spectrum sharing is confined within the borders of a single nation (e.g., TNs). However, the extensive operational scope of NTNs necessitates the implementation of robust regulatory mechanisms to uphold the integrity of spectrum agreements and preempt potential disruptions.


\subsubsection*{Security}

If spectrum sharing is not employed, the risks of security threats and attacks may be more contained and readily discernible within their respective network segments. However, when spectrum sharing is introduced, the demarcation between system boundaries becomes blurred. It becomes more challenging to ascertain whether anomalies in network behavior are the result of legitimate spectrum sharing activities or potential security threats. This ambiguity is exacerbated by the increased openness inherent in spectrum sharing, which can augment the attack surface for potential adversaries. Consequently, while O-RAN based spectrum sharing presents numerous advantages in terms of connectivity and resource utilization, it simultaneously necessitates a comprehensive and adaptive security framework to address these new and complex challenges. 

\section{Conclusions}   \label{S:Conc}

Addressing the challenges of spectrum scarcity and flexible allocation in NTNs is imperative, prompting a proactive exploration of spectrum-sharing strategies for future networks. This article has examined the effectiveness of spectrum sharing through marketplaces tailored for O-RAN based NTNs and TNs, elucidating potential use cases. As satellite IoT services aim to cover vast territories, including open seas, the efficient utilization of spectrum becomes pivotal. Notably, the application of spectrum sharing proves particularly pertinent for IoT networks operating over extensive regions. The role of spectrum marketplaces stands out as a crucial mechanism, facilitating the trade of under-utilized spectrum resources across multiple countries. This balanced approach ensures that the challenges of spectrum allocation are met with innovative solutions, paving the way for enhanced connectivity and efficiency in both NTNs and NTs.

\bibliographystyle{ieeetr}
\bibliography{si}

\end{document}